\documentclass[sn-mathphys,Numbered ]{sn-jnl}

\usepackage{graphicx}%
\usepackage{multirow}%
\usepackage{amsmath,amssymb,amsfonts}%
\usepackage{amsthm}%
\usepackage{mathrsfs}%
\usepackage[title]{appendix}%
\usepackage{xcolor}%
\usepackage{textcomp}%
\usepackage{manyfoot}%
\usepackage{booktabs}%
\usepackage{algorithm}%
\usepackage{algorithmicx}%
\usepackage{algpseudocode}%
\usepackage{listings}%

\theoremstyle{thmstyleone}%
\theoremstyle{thmstyletwo}%

\theoremstyle{thmstylethree}%

\begin{document}
	
	\title[Article Title]{ Charge fluctuations in charge regulated systems: dependence on statistical ensemble}
	
	
	\author[1]{\fnm{Amin} \sur{Bakhshandeh}}\email{bakhshandeh.amin@gmail.com}
	
	\author[1]{\fnm{Yan} \sur{Levin}}\email{levin@ufrgs.br}

	\affil[1]{\orgdiv{Instituto de F\'isica}, \orgname{Universidade Federal do Rio Grande do Sul}, \orgaddress{ \postcode{ 91501-970}, \state{Porto Alegre,RS}, \country{Brazil}}}

	
	\abstract{We investigate charge regulation of nanoparticles in concentrated suspensions, focusing on the effect of different statistical ensembles. We find that the choice of ensemble does not affect the mean charge of nanoparticles, but significantly alters the magnitude of its fluctuation.  Specifically, we compared the behaviors of colloidal charge fluctuations in
		the semi-grand canonical and canonical ensembles, and identified significant differences between the two. The choice of ensemble -- whether the system is isolated or is in contact with a reservoir of acid and salt -- will, therefore, affect the  Kirkwood-Shumaker fluctuation-induced force inside concentrated suspensions. Our results emphasize the importance of selecting
		an appropriate ensemble that accurately reflects the experimental conditions
		when studying fluctuation-induced forces between polyelectrolytes, proteins, and
		colloidal particles in concentrated suspensions.}

	
	
	
	\maketitle
	
	\section{Introduction}\label{sec1}
	
	It is a great pleasure to contribute this paper to the special issue of EPJE that honors many contributions of Philip (Fyl) Pincus in the field of Soft Matter Physics~\cite{PhysRevLett.101.188101,PhysRevLett.101.208305,Microsurface,Tamashiro,Pincus_1998,Tadmor,MacKintosh_1990,gelbart2000dna}.  Much of Fyl's work has been focused on understanding complicated 
	effects resulting from Coulomb force in condensed matter systems.  We hope that Fyl will find the following paper of interest.   
	
	Electrostatic interactions are ubiquitous in physics, chemistry, and biology~\cite{smith2006electrostatic,honig1986electrostatic,venkataraman2017emphasizing,PhysRevLett.101.128101,Naydenov,andelman2006introduction,bakhshandeh2018theoretical,C9SM02532D,pharmaceutics13091365,ELOUAHABI2005336,Vijayanathan}. 
	They play a vital role in the stability and function of biological molecules~\cite{zhong2012colloidal} and the structure and dynamics of electrolyte solutions. The magnitude of the force between
	charged particles is heavily dependent on the dielectric constant of the solvent. In fact, water is essential for many biological processes~\cite{margulis2000life,Ball2013,israelachvili1992intermolecular}, and its unique properties largely stem from its strong electrostatic interaction with the charged solutes. Due to the high dielectric constant of water~\cite{malmberg1956dielectric,owen1961dielectric,Collie_1948,eisenberg2005structure,Binazadeh,Howerton,cheng2004size,davidchack2012ice,luo2012electrostatic,sidhu1999effect,hurd1985electrostatic}, electrostatic interactions between charged atoms and molecules are greatly reduced, resulting in a favorable solvation free energy. The high dielectric constant of water also causes the dissociation of molecules into ions~\cite{booth1951dielectric}, resulting in electrolyte solutions. 
	The formation of hydrogen bonds between water and biomolecules helps to stabilize their structures and facilitate their interactions with other molecules in biological cells. 
	
	In colloidal science, the charging process of colloidal particles in an electrolyte solution is called charge regulation~\cite{ninham1971electrostatic,frydel2019general,bakhshandeh2019charge,bkh22,bakhshandeh2022reactive,bakhshandeh2021equilibrium,bakhshandeh2020charge,avni2018charge,tagliazucchi2010self,lund2013charge,longo2011molecular,heinen2014coupling,lund2005charge,boon2011charge}. The main reason for this phenomenon is that macromolecules, in general, contain functional acidic and basic groups that can become protonated or deprotonated, depending on the ionic strength, pH, and solute volume fraction. In equilibrium, the charge of macromolecules will fluctuate around some mean value.  The correlated fluctuations of charge can result in an effective attractive Kirkwood-Shumaker (KS) force between two like-charged macromolecules close to their isoelectric point~\cite{kirkwood1952forces,kirkwood1952influence}.  A theoretical prediction, which was confirmed by Timasheff {\it{et. al.}}~\cite{timasheff1955studies} using light-scattering techniques.    
	
	The Kirkwood-Shumaker model has been successfully applied in many areas of biophysics, such as protein-protein interactions, protein aggregation, and protein crystallization. In particular, it has been used to explain the phenomenon of liquid-liquid phase separation in protein solutions~\cite{bozic2021site}. The KS interaction is also relevant for understanding the behavior of polyelectrolytes in solutions, for which the long-range Coulomb interaction between charged macromolecules plays a crucial role~\cite{ avni2019charge}.
	The KS model, however, has limitations, particularly when applied to systems with highly charged macromolecules or in the presence of multivalent ions. In such cases, other factors, such as ion correlation effects, must be taken into account~\cite{levin2002electrostatic}. 
	Despite these limitations, the KS model remains a valuable tool for understanding the long-range interactions between macromolecules in solutions. It provides a useful framework for interpreting experimental data and can guide the development of new theoretical models.  Since the KS force  arises from correlated fluctuations of the macromolecular charge,  it is of fundamental importance to understand how much this charge fluctuates around its
	mean value for different experimental conditions.   
	
	Most experimental systems consist of an isolated suspension at some fixed volume fraction of solute.  On the other hand,
	to perform simulations one often uses a semi-grand canonical approach, in which suspension is effectively separated from the reservoir of acid and salt by a semi-permeable membrane, which allows for a free exchange of ions, but prevents the passage of macromolecules~\cite{labbez2007new,bakhshandeh2022reactive, curk2021charge,longo2011molecular}.  It is intuitive that in such open systems, macromolecular charge will fluctuate more than in closed (isolated) systems.  The goal of the present paper is to quantify this difference.
	
	


	\section{Non-interacting systems}~\label{s1}
	
	We start by studying an ideal system in which electrostatic interactions are turned off.  
	The model consists of a  nanoparticle of radius $a$ and $Z$ (negative) active surface sites.  Each active site can adsorb one proton to become protonated.  A nanoparticle is placed at the center of a spherical cell of radius $R$, such that $\eta=a^3/R^3$, where $\eta$ is the volume fraction of suspension.  The cell also contains noninteracting point ions: Na$^+$, Cl$^-$ and H$^+$.  The proton H$^+$ can associate with an adsorption site, resulting in a free energy gain of $-\ln(\text{K}_{eq}/\Lambda^3)$,  where the equilibrium constant $\text{K}_{eq}$ is the partition partition function of a bound state. In this ideal model, the ionic charge is used just to distinguish cations from anions and to preserve the overall charge neutrality inside the cell.
	
	\subsection{Canonical Theory}\label{ct}
	The free energy of protons inside an isolated (canonical) cell can be written as,
	\begin{equation}\label{e2}
		\begin{aligned}
			\beta \mathcal{F}(n)  = -n\ln\left( \frac{\text{K}_{eq} }{\Lambda^3_{\text{H}^+}}\right) +n\ln\left( \frac{n }{Z}\right)+\left(Z-n\right)\ln\left( 1-\frac{n }{Z}\right) +\\\left(N_t-n\right)\ln\left(\frac{\Lambda^3_{\text{H}^+}\left(N_t-n\right)}{V}\right)-\left(N_t-n\right),
		\end{aligned}
	\end{equation}
	where $N_t$ is total number of protons inside the system, of which $n$ are in a bound state with the surface groups.  $V$ is the free volume of the cell and $\text{K}_{eq}$ and $\Lambda_{\text{H}^+}$ are the equilibrium constant and the de Broglie thermal wavelength, respectively.  The first term in the expression above is the chemical energy of association between proton and an active site.  Second and third terms are the entropic contributions of the bound protons, while the last two terms are the entropic contributions of free protons.
	Minimizing Eq.~\ref{e2} with respect to $n$, we obtain the equilibrium (average) number of protonated sites:
	\begin{equation}\label{e3}
		n^*= 	\frac{1}{2}
		\left(\frac{{V}}{{\text{K}_{eq}}}+{N_t}+{Z}-\sqrt{\left(\frac{{V}}{{\text{K}_{eq}}}+{N_t}+
			\text{Z}\right)^2-4 {N_t}
			\text{Z}}
		\right).
	\end{equation}
	It is important to stress that the number of associated protons is not fixed, but fluctuates around the value $n^*$, with the average colloidal charge given by $Q^*=-(Z-n^*) q$, where $q$ is the proton charge. 
	The charge fluctuation is characterized by $\langle (\Delta Q)^2 \rangle = \langle Q^2 \rangle- \langle Q \rangle^2=q^2 (\langle n^2 \rangle- \langle n \rangle^2)$.
	The probability that $n$ active sites are protonated is proportional to $\mathrm{e}^{-\beta\mathcal{F}(n)}$. Expanding the free energy $\mathcal{F}(n)$ around the equilibrium $n^*$ up to second order and using the saddle point approximation, 
	the fluctuation of charge in canonical ensemble is determined to be: 
	\begin{equation}\label{e4}
		\sigma_t^2 \equiv\frac{\langle (\Delta Q)^2 \rangle}{q^2}=\frac{ n^* \left(n^* - N_t \right) \left(n^* - Z\right)}{N_t Z -n^{*2}  }.
	\end{equation}

	\subsection{Semi-grand Canonical Theory}\label{gct}
	Titration simulations are usually performed in a semi-grand canonical ensemble in which microions are free to exchange with the reservoir, while nanoparticles are confined to stay within the system~\cite{bakhshandeh2022reactive}.  In a real experimental system such setup requires a semipermeable membrane that separates system from the reservoir of acid and salt. Since the counterions are free to diffuse into reservoir, their efflux will result in an electric field across the membrane that will oppose the flow.    The concentration of ions inside the system will, in general, be different from the concentrations in the reservoir.  When performing semi-grand canonical simulations, it is important to keep in mind that the simulation cell is at a different electrostatic potential than the reservoir.  This electrostatic potential difference is known as the Donnan potential.   Often, Donnan potential is implicitly taken into account when performing semi-grand canonical simulations of charged system by forcing the insertion and deletion moves to be done in cation-anion pairs, which effectively cancels the Donnan potential.  However, presence of the Donnan potential is often neglected when performing constant pH titration simulations~\cite{C8SM02085J,D2SM01393B}.  
	
	Neglecting the Coulomb and steric interactions between the ions, the ideal partition function for a system containing a nanoparticle inside a spherical cell connected to a reservoir of acid and salt at concentrations $C_H$ and $C_s$, respectively, can be written as:
	\begin{equation}\label{er}
		\begin{aligned}
			\Xi = \sum_{N_{\text{Na}},N_{\text{H}}N_{\text{Cl}}}^\infty\sum_{n=0}^{Z}\frac{Z!}{n!(Z-n)!}\exp\left[-\beta\left(-(Z-n)q \phi-n\ln\frac{\text{K}_{eq}}{\Lambda^3}-n\mu_{\text{H}}\right)  \right]\\ \frac{1}{N_{\text{Na}}!}\frac{1}{N_{\text{Cl}}!}\frac{1}{N_{\text{H}}!}\left(\frac{V\mathrm{e}^{-\beta(q\phi-\mu_{\text{Na}})}}{\Lambda^3}\right)^{N_{\text{Na}}}\left(\frac{V\mathrm{e}^{\beta(q\phi+\mu_{\text{Cl}})}}{\Lambda^3}\right)^{N_{\text{Cl}}}\left(\frac{V\mathrm{e}^{-\beta(q\phi-\mu_{\text{H}})}}{\Lambda^3}\right)^{N_{\text{H}}},
		\end{aligned}
	\end{equation}
	where $N_{\text{H}}$, $N_{\text{Cl}}$, $N_{\text{Na}}$ are the number of particle of specie H, Cl, Na inside the system;  and V, $\mu_{\text{H}}$, $\mu_{\text{Na}}$, $\mu_{\text{Cl}}$, $\Lambda$ are the free volume, chemical potential of H, Na, Cl and the thermal de Broglie wavelength, respectively. In general, we allow the system to be at a different potential from the reservoir, this accounts for the presence of the electrostatic potential $\phi$ in the partition function.   
	The two summations in Eq.\ref{er} are decoupled and can be performed separately. 
	The partition function for protonation of surface sites reduces to:
	\begin{equation}\label{key}
		\sum_{n=0}^{Z}\frac{Z!}{n!(Z-n)!}\exp\left[-\beta\left(-(Z-n)q\phi-n\ln\frac{\text{K}_{eq}}{\Lambda^3}-n\mu_\text{H}\right)  \right] = 
		\left(\mathrm{e}^{\beta q\phi}+\mathrm{e}^{ \ln \frac{\text{K}_{eq}}{\Lambda^3}+ \mu_\text{H}}\right)^Z,
	\end{equation}
	while the partition function for free ions is: 
	\begin{equation}\label{e7}
		\begin{aligned}
			\sum_{N_\text{H},N_\text{Na},N_\text{Cl}=0}^\infty\frac{1}{N_{\text{Cl}}!}\frac{1}{N_{\text{H}}!}	\frac{1}{N_{\text{Na}}!}\left({C_{\text{Na}}V\mathrm{e}^{-\beta q \phi}}\right)^{N_{\text{Na}}}\left({C_{\text{Cl}}V\mathrm{e}^{ \beta q \phi}}\right)^{N_{\text{Cl}}}\left({C_{\text{H}}V\mathrm{e}^{-\beta q \phi}}\right)^{N_{\text{H}}}=\\
			\exp\left({C_{\text{H}}V\mathrm{e}^{-\beta q\phi}}\right)	\exp\left({C_{\text{Na}}V\mathrm{e}^{-\beta q\phi}}\right)	\exp\left({C_{\text{Cl}}V\mathrm{e}^{ \beta q\phi}}\right),
		\end{aligned}
	\end{equation}
	where we have used $\beta \mu_i=\ln \left(C_i \Lambda^3\right)$ for ideal chemical potential inside the reservoir.
	
	The net charge inside the cell is:
	\begin{equation}\label{e8}
		\langle   \mathcal{Q}_{net}  \rangle  = -\frac{1}{\beta} \frac{d\ln\Xi}{d\phi}= -\frac{ \mathrm{e}^{\beta\phi}  {Z q}}{ {C}_\text{H}
			\text{K}_{eq}+ \mathrm{e}^{\beta\phi}} + q \left( {C}_\text{H} + {C}_{\text{Na}} - {C}_{\text{Cl}} \mathrm{e}^{2\beta\phi}  \right) V \mathrm{e}^{-\beta\phi}.
	\end{equation}
	In the thermodynamic limit, the cell must be charge neutral, which means that $	\langle  \mathcal{Q}_{net} \rangle=0$.  This condition determines the Donnan potential $\phi_d$.  The charge of a nanoparticle at equilibrium in a semi-grand canonical system will then be:
	\begin{equation}\label{e8a}
		\langle   Q  \rangle  = -\frac{ \mathrm{e}^{\beta\phi_d}  {Z q}}{ {C}_\text{H}
			\text{K}_{eq}+ \mathrm{e}^{\beta\phi_d}}.
	\end{equation}
	
	The fluctuation in the net charge inside the system can be obtained from the second derivative of the partition function with respect to $\phi$.  These fluctuations decouple into those due to fluctuation of colloidal charge and of free ions in the bulk.  The fluctuation in the charge of the nanoparticle is calculated to be:
	\begin{equation}\label{key}
		\sigma_t^2\equiv\frac{\langle (\Delta Q)^2 \rangle}{q^2}=-\frac{1}{q^2 \beta}\frac{d \langle Q \rangle}{d\phi_d} = \frac{C_\text{H}  \text{K}_{eq} Z \mathrm{e}^{\beta q \phi_d}}{\left(\mathrm{e}^{\beta q \phi_d} + C_\text{H} \text{K}_{eq}\right)^2}.
	\end{equation}
	
	Comparing the charge fluctuations in canonical and semi-grand canonical systems, we see that the fluctuations of colloidal charge in the semi-grandcanonical system are much larger than in a canonical system. To compare the two ensembles, in the canonical system we put exactly the same number of protons and ions into the cell as the averages obtained in the semi-grand canonical system.   See, for example, the values of $\sigma_t^2$ in Tables 1 and 3, respectively for the two systems. To go beyond the ideal models, requires Monte Carlo simulations methods.
	Below, we briefly present the semi-grand canonical~\cite{labbez2007new,bakhshandeh2022reactive},  and canonical simulations methods that can be used to explore charge regulation and fluctuations in the two ensembles~\cite{johnson1994reactive,D2SM01393B}.

	\section{Titration Algorithms}
	
	When performing canonical simulations, the system can only exchange heat with the surrounding environment, while the number of protons and other ions is conserved inside the simulations cell.  On the other hand in semi-grand canonical simulations, the ions and protons can be exchanged with the reservoir, and the average concentrations of ions inside the simulation cell are determined by the thermodynamic equilibrium -- equivalence of electrochemical potentials in the system and the reservoir.  Furthermore, as was discussed previously, the simulation cell is at a different electrostatic potential from the reservoir.  The potential difference between the cell and the reservoir is the Donnan potential.  In systems with finite volume fractions of nanoparticles, proteins, or polyelectrolytes, this potential difference can not be ignored and must be taken into account when performing simulations.  
	
	\subsection{ Semi-grand canonical Monte Carlo method }
	
	In semi-grand canonical Monte Carlo (sGCMC) simulations we need to perform protonation/deprotonation moves  as well as  insertion/deletion  moves into/from the cell.  Since the simulation cell is at a different electrostatic potential than the reservoir, upon entering the cell an ion acquires an additional electrostatic energy $q_i\phi_D$, where $\phi_D$ is the Donnan potential. Taking this into account  the usual   grand-canonical acceptance probabilities  for addition and removal of ions are modified to:  
	\begin{equation}\label{eqd1}
		\begin{split}
			\phi_{add} = \min\left[1, \frac{V c_i}{N_i +1}\mathrm{e}^{- \beta\left( \Delta U- \mu_{ex}+ q_i\varphi_D  \right) }\right],\\
			\phi_{rem} = \min\left[1, \frac{ N_i}{V c_i}\mathrm{e}^{- \beta\left( \Delta U+ \mu_{ex}- q_i \varphi_D  \right) }\right].
		\end{split}
	\end{equation}
	For reaction moves,  proton can enter from the reservoir and react with an adsorption site resulting in its protonation.   Alternatively a protonated site can become deporonated, with the proton moving to reservoir.  Again, when a proton moves into or out of the system, the Donnan potential must be taken into account.  The acceptance probabilities for protonation and deprotonation moves can then be written as:
	\begin{equation}\label{eqRM}
		\begin{split}
			\phi_p =\min\left[1, \mathrm{e}^{-\beta \left(\Delta U  +  \Delta F_p +q\varphi_D   \right)}\right],\\
			\phi_d =\min\left[1,\mathrm{e}^{-\beta \left(\Delta U   + \Delta F_d  -q\varphi_D     \right)}\right],
		\end{split}
	\end{equation}
	where  $\beta \Delta F_p=-\ln(\text{K}_{eq}/\Lambda^3_{\text{H}})-\mu_{\text{H}}$ is the chemical free energy change due to removal of proton from the reservoir and its reaction with an isolated adsorption group.  The chemical potential of proton in reservoir is $ \beta \mu_{\text{H}}=\ln(c_{\text{H}}\Lambda^3_{\text{H}})+ \beta \mu_{ex}$, where $\mu_{ex}$ is the excess chemical potential of ions in the reservoir.  The deprotonation energy is then $\Delta F_d=-\Delta F_p$.  
	Since the Donnan potential is not known {\it a priori}, it is convenient to perform insertion/deletions moves using cation-anion pairs~\cite{labbez2007new,bakhshandeh2022reactive}.  This way the Donnan potential cancels from the acceptance probabilities.  Similarly, a protonation/deprotonation move can be combined with an insertion/deletion of an anion into the the cell.  
	The acceptance probabilities for such pair protonation/deprotonation moves become:
	\begin{equation}\label{eqRM2}
		\begin{split}
			\bar \phi_p = \min\left[1, \frac{ c_{\text{H}}\text{K}_{eq} V c_{\text{Cl}}}{(N_{\text{Cl} } +1)}\mathrm{e}^{-\beta \left(\Delta U -2\mu_{ex}        \right)}\right],\\
			\bar\phi_d =\min\left[1,  \frac{ N_{\text{Cl}}}{c_{\text{H}}\text{K}_{eq}V c_{\text{Cl-}}}\mathrm{e}^{-\beta \left(\Delta U +2\mu_{ex}      \right)}\right], 			
		\end{split}
	\end{equation}
	where $N_{\text{Cl} }$ is the number of  anions inside the cell, $V$ is the free volume, $\Delta U$ is the difference of energy for the pair move.  For simplicity, here we consider that all ions are monovalent and are hard spheres of the same radius, so that  $\mu_{ex}$ is the same for all the ions.  
	
	The algorithm described above can be applied to any system in which reactions take place.  We start by studying an ideal non-interacting 
	system described in Section \ref{gct}.   In the Table 1, we compare the results of simulations with the theory for a nanoparticle  of radius  80~\AA\  with $Z=600$ surface active groups, inside a simulation cell of radius $R=150$\AA.   The equilibrium constant for sites is taken to be $K_{eq}=1216092$~\AA$^3$.  The reservoir contains acid at concentration $C_H$ and salt at concentration $C_s=10$~mM.  The mean charge of a nanoparticle and its fluctuation calculated using both theory and simulations are presented in Table 1.  
	\begin{table}[h]
		\caption{ Nanoparticle charge $Q$ and its fluctuation $\sigma^2$ obtained from simulations (s)  and theory (t) for different concentrations of acid in the reservoir $C_\text{H}$ The concentration of salt in the reservoir is fixed at $C_s=10$~mM. The average number of ions of each type present inside the simulation cell after equilibration is also provided.}\label{tab1}

		\begin{tabular}{c||ccccccc}
			\toprule%
			C$_\text{H}$[M]	  &  $Q^s/q$ &  $Q^t/q$  &  $\sigma_s^2$ &$\sigma_t^2$&$N_\text{H}$&$N_\text{Na}$&$N_\text{Cl}$ \\
			\hline
			$10^{-3}$ 	& $-195.4$& $-195.7$&88.3&131.8&424 &203 &28\\
			\hline
			$10^{-4}$ 	& $-417.9$&$-418.7$ &98.9&126.4&185 &426 &12\\	 
			\botrule
		\end{tabular}
		
	\end{table}
	We see that while the effective charge is in perfect agreement with simulations, the fluctuations show significant deviations.  The reason for this is that the pair insertion moves restrict the charge fluctuations inside the cell.  Indeed, if we perform simulations using individual insertions, Eq. (\ref{eqRM}) with the Donnan potential fixed at the value predicted by the theory, we obtain exactly the same numbers of ions inside the simulation cell and the same colloidal charge as found using the pair insertion algorithm, Eq. (\ref{eqRM2}). Thus, as expected, the Donnan potential leads to an  overall charge neutrality {\it on average}.  Furthermore,
	individual insertion algorithm with Donnan potential, results in colloidal charge fluctuations very similar to the ones predicted by the theory, see Table~\ref{T2}.
	\begin{table}[h]
		\caption{ Colloidal charge and its fluctuations obtained using individual insertion algorithm (o) with the Donnan potential fixed at the value predicted by the theory (t). Compare the fluctuations obtained using the individual insertions algorithm $\sigma_{o}^2$, with the ones obtained using the pair insertion method $\sigma_{s}^2$ presented in Table 1.  All other parameters are the same as in the Table 1.}\label{T2}

		\begin{tabular}{c||cccc}
			\toprule%
			C$_\text{H}$[M]	  &  $Q^o/q$ &  $Q^t/q$  &  $\sigma_t^2$ &$\sigma_{o}^2$ \\
			\hline
			$10^{-3}$ 	& $-196.1$& $-195.7$&131.8& 135.4\\
			\hline
			$10^{-4}$ 	& $-418.8$&$-418.7$ &126.4&131.0\\	 
			\botrule
		\end{tabular}

	\end{table}
	Clearly, restricting the insertion moves to keep the system charge neutral at each Monte Carlo step, strongly affects the fluctuations of colloidal charge.  With these insights, we are now ready to explore isolated canonical systems.  
	
	\subsection{ Canonical reactive Monte Carlo method  }
\begin{figure}[H]
		\centering
		\includegraphics[width=0.7\linewidth]{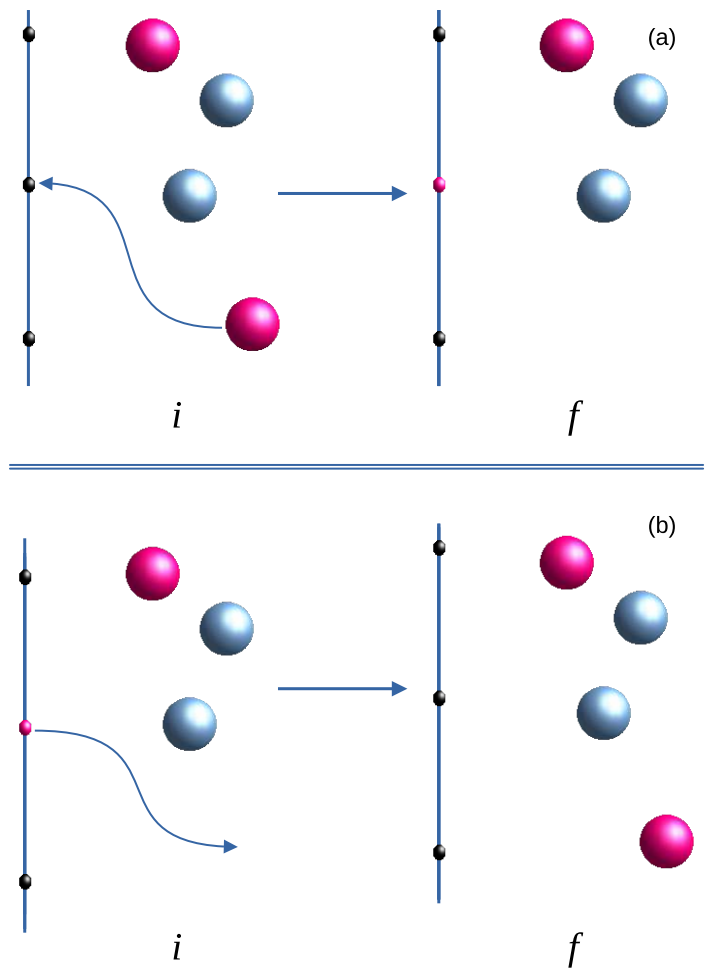}
		\caption{(a) Protonation and (b) deprotonation moves in canonical ensemble:  ($i$)   initial state,($f$)  final state.}
		\label{fig1}
	\end{figure}

	When performing a canonical simulation, the number of protons and ions inside the simulation cell is fixed.  However, the protons can either be in a bound state or free, see Fig.~\ref{fig1}. The average number of bound protons will determine the equilibrium charge of nanonparticles.  
	$$$$$$$$$$$$$$$$$$$$  
	In a protonation move, a proton that is initially in the bulk, moves to the adsorption site.   The probability for initial and final configuration are then proportional to:
	\begin{eqnarray}\label{eq92} 
		&&\Pi_i \sim\frac{V^N}{\Lambda^{3N}N!}\mathrm{e}^{- \beta U_N}, \nonumber \\
		&&\Pi_f\sim \frac{V^{N-1}}{\Lambda^{3(N-1)}\left(N-1\right)!}\mathrm{e}^{- \beta U_{N-1}+\ln \text{K}_{eq}/\Lambda^3},   
	\end{eqnarray}
	where $N$ is the number of free protons and $U_N$ is the total electrostatic energy of the system.  Similarly for 
	deprotonation move
	\begin{eqnarray}\label{eq91} 
		&&\Pi_i \sim \frac{V^N}{\Lambda^{3N}N!}\mathrm{e}^{-\beta U_N }, \nonumber \\
		&&\Pi_f \sim \frac{V^{N+1}}{\Lambda^{3(N+1)}\left(N+1\right)!}\mathrm{e}^{- \beta U_{N+1}-\ln \text{K}_{eq}/\Lambda^3},   
	\end{eqnarray}
	Using the usual detailed balance argument, the acceptance probabilities for deprotonation and protonation moves can now be written as:
	\begin{eqnarray}\label{eq9} 
		&&\text{P}_d=\min\left[1,\frac{ V  }{\text{K}_{eq}\left(N_{\text{H}^+} +1\right)} e^{-\beta \Delta U} \right], \nonumber \\
		&&\text{P}_p =\min\left[1,\frac{\text{K}_{eq}N_{\text{H}^+}} { V  } e^{-\beta \Delta U}\right]  ,
	\end{eqnarray}
	
	We can now check the consistency of the two simulation methods.  We first run the semi-grand canonical simulation to determine the number of ions, free protons, and the nonoparticle charge inside the simulation cell for a given concentration of acid and salt in the reservoir. We can then strip all the associated protons and put them into the bulk of the simulation cell, so that all sites are again deprotonated.  We then run canonical reactive Monte Carlo algorithm, Eq. (\ref{eq9}) to determine the equilibrium number of protonated sites and the equilibrium colloidal charge.  The data is presented in Table 3.  While the charge of the nanoparticle calculated using canonical simulation is in perfect agreement with the results of the semi-grand canonical algorithm, the fluctuation in the charge is significantly lower than what is observed in an open system.  Furthermore, we see that for a canonical ideal system, both the nanoparticle charge and its fluctuation are in excellent agreement with the predictions of the theory.   With the insights gained from studying non-interacting ideal systems, we are now ready  to explore charge fluctuations in non-ideal systems, with ions of finite size interacting through Coulomb potential.    
	\begin{table}[h]
		\caption{Colloidal charge and its fluctuations for ideal system calculated using canonical simulations (s) compared with the predictions of theory (t), Eqs.(\ref{e3}) and (\ref{e4}).  The number of ions and protons inside the simulation cell is exactly the same as are the averages found using sGCMC simulations, see Table~\ref{tab1}}.

		\begin{tabular}{ccc||cccc}
			\toprule%
			$N_\text{H}$&$N_\text{Na}$&$N_\text{Cl}$ & $Q^s/q$ &  $Q^t/q$ &   $\sigma_s^2$  &  $\sigma_t^2$ \\   
			\hline
			424 &203 &28& $-195.4$	&$-195.5$&17.4&17.6\\
			\hline
			185 &426 &12&  $-418.6$	&$-418.6$&4.2&4.1	 \\

			\botrule
		\end{tabular}

	\end{table}
	
		

	
	

	\section{Systems with Coulomb interactions}~\label{s2}
	
	We now consider a nanoparticle of radius of 80~\AA\ with $Z=600$ active negative point sites distributed uniformly on its surface~\cite{bkh22}, inside a spherical cell of radius $R=150$~\AA.
	The intrinsic pK$_a$ of surface groups is set to pK$_a \equiv -\log_{10}$K$_a$=5.4, corresponding to carboxylic acid.  Recall that acid dissociation constant is $\text{K}_a=1/\text{K}_{eq}$. All ions are modeled as hard spheres of radius $r=2$~\AA, with point charge at the center.  Water is treated as a uniform medium of dielectric constant $\epsilon=78$. The electrostatic energy now includes Coulomb ion-ion, ion-site, and site-site interactions.  Again, we first run the sGCMC simulations to determine the average number of ions, the number of free protons,  the mean charge of the nanoparticle, and its fluctuation $\sigma^2$, for a given concentration of acid and salt in the reservoir, Table 4.  
	It is interesting to compare the results obtained using the sGCMC simulations with the recently developed theory that allows us to accurately predict the effective colloidal charge in concentrated suspensions~\cite{bkh22}.  
	
	The effective charge (number of deprotonated groups) predicted by the theory is~\cite{bkh22}:
	\begin{equation} \label{zeff2}
		Q^t=-\frac{Z q}{1+ \text{K}_{eq} c_a e^{-\beta(q \phi_0-\phi_{disc}-\mu^{ex}-\mu_{sol})}},
	\end{equation} 
	where $ \mu_{sol}$ the  electrostatic solvation free energy of a charged site:
	\begin{equation}\label{sol}
		\beta \mu_{sol}=\frac{\lambda_B}{2}\int_0^\infty\frac{k-\sqrt{\kappa^2 +k^2}}{k+\sqrt{\kappa^2 +k^2}} e^{-2 k r_{ion}}dk,\ 
	\end{equation}
	and $\phi_0$ is the mean-field electrostatic potential at the surface determined from the solution on non-linear Poisson-Boltzmann equation.   $\phi_{disc}$ is the  correction due to the discreteness of surface groups:
	\begin{equation}\label{EOCP2}
		\beta \phi_{disc}    = -\frac{\lambda_b M Q^t}{q a \epsilon_w \sqrt{Z}}, 
	\end{equation}
	where $M$ is the Madelung constant for hexagonal crystal state of the one component plasma~\cite{bkh22}. The excess chemical potential of ions in the reservoir  $\mu_{ex} $ can be approximated as the sum of the  mean spherical approximation (MSA) chemical potential and the Carnahan-Starling expression for the excluded volume 
	contribution~\cite{carnahan1969equation,carnahan1970thermodynamic,adams1974chemical,MACIEL2018,ho1988mean,ho2003interfacial,levin1996criticality,waisman1972mean,blum1975mean}, which are:
	\begin{equation}
		\beta \mu_{MSA} = \frac{\lambda_B\left( \sqrt{1+2 \kappa d}-\kappa d -1\right)} {d^2\kappa},
		\qquad
		\beta\mu_{CS} = \frac{8\eta-9 \eta^2+3\eta^3}{\left(1-\eta\right)^3},
		\qquad
	\end{equation}
	where  $\eta=\frac{\pi d^3}{3} c_t$,  $d$ is the ionic diameter,   $c_t=c_s+c_a$ is the total concentration of salt and acid, $\lambda_B=q^2/\epsilon_w k_B T$ is the Bjerrum length, and  $\kappa= \sqrt{8 \pi \lambda_B c_t}$ is the inverse Debye length. 
	
	At the same level of approximation, the fluctuation of surface charge can be written as:
	\begin{equation} \label{zeff3}
		\sigma_t^2=\frac{Z~ \text{K}_{eq} c_a \mathrm{e}^{\left(-\beta[\phi_0-\phi_{disc}-\mu^{ex}-\mu_{sol}]\right)} }{\left(1+ \text{K}_{eq} c_a  \mathrm{e}^{\left(-\beta[\phi_0-\phi_{disc}-\mu^{ex}-\mu_{sol}]\right)}\right)^2},
	\end{equation} 
	Comparison of theory with the sGCMC results are shown in Table 4.  We see that the theory again agrees well with the colloidal charge calculated in simulations, but the fluctuations in the nanoparticle charge differ significantly from the predictions of the theory.  This is similar to what was found in the ideal case when comparing theory with sGCMC simulations that used pair insertions. The pair insertion algorithm restricts charge fluctuations inside the simulation cell, affecting also the fluctuations of nanoparticle charge.

	\begin{table}[h]
		\caption{Colloidal charge  and its fluctuations for interacting system: theory (t) and simulations (s).  The Table also presents the average number of ions and free protons inside the cell.  The same numbers are used to perform canonical simulations, results of which are presented in the Table 5.}\label{tab3}

		\begin{tabular}{c||ccccccc}
			\toprule%
			$C$ M&$Q^s/q$	&  $\sigma_s^2$ &  $N_{\text{H}}$$^+$&$N_{\text{Na}}$$^+$&$N_{\text{Cl}}$$^-$&$ Q^t/q$&$\sigma^2_t$\\
			\hline
			$10^{-5}$	&	$-66.7	$&  29.0  &  533&112&45&$-65.4$ &58.2\\
			\hline
			$10^{-5.5}$	&	$-106.6 $	 & 39.4&493&145&39&  $-102.9$ &85.1\\
			\hline
			$10^{-6}$	&	$-160.6 $	&  52.4 &439&195&34& $-152.9$ &113.8\\
			\botrule
		\end{tabular}

	\end{table}
	
	We next run the canonical reactive Monte Carlo algorithm, Eq. (\ref{eq9}).  The simulation cell contains the same number of ions and protons as was obtained using sGCMC simulations previously.  After the simulation reaches equilibrium,  we see that the average charge of a nanoparticle calculated using canonical simulation is in perfect agreement with the results of the semi-grand canonical algorithm.  On the other hand, canonical fluctuations of the nanoparticle charge are almost 4 orders of magnitude lower than what was found for an open system, see Table 5. 
	
	\begin{table}[h]
		\caption{ $Q$ and its fluctuations in the canonical ensembles for an interacting system.  Note that the fluctuations are four orders of magnitude smaller than observed in the sGCMC case for exactly the same parameters, compare with Table 4.}\label{tab4}

		\begin{tabular}{ccc||cc}
			\toprule%
			$N_{\text{H}}$$^+$&$N_{\text{Na}}$$^+$&$N_{\text{Cl}}$$^-$ & $Q^s/q$    & $\sigma^2$ \\
			\hline
			533	   &  112 & 45&$-67.11$ &0.11  \\
			\hline
			493&145&39&  $-106.04 $   &0.04     \\
			\hline
			439&195&34&	  $-161.01 $ &0.01  \\
			\botrule
		\end{tabular}

	\end{table}
	
	\section{Conclusions}\label{sec13}~\label{s3}
	
	In this paper we investigated charge regulation in isolated (canonical) and open (semi-grand canonical) systems.  An open system can exchange heat, ions, and protons with an external reservoir, while a closed system can only exchange heat.  In both cases, equivalence of ensembles extends to the prediction of the effective charge of nanoparticles -- both ensembles predict exactly the same charge.  On the other hand, the fluctuations of colloidal charge are very different, with canonical fluctuations 4 orders of magnitude smaller than the semi-grand canonical ones. Since the Kirkwood-Shumaker force depends on charge fluctuations, its manifestation should  be very different in the two ensembles, in particular for concentrated suspensions.  At infinite dilution, we expect the difference between the two ensembles to vanish.  In the future work, we will use the simulation methods discussed in this paper to explicitly calculate Kirkwood-Shumaker force between nanoparticles in different ensembles.

	\bmhead{Acknowledgments}
	
	This work was partially supported by the CNPq, the
	CAPES, and the National Institute of Science and Tech-
	nology Complex Fluids INCT-FCx.

	\bibliography{ref}
	
\end{document}